# Frequency Response of the Mechanochemically Synthesized *AgI-Ag₂O-B₂O₃* Superionic Glasses

**Ankur Verma and K. Shahi** [a]

Solid State Ionics Laboratory, Advanced Center for Materials Science

[a] also in the Department of Physics

Indian Institute of Technology Kanpur, India–208016

Email: ankurv@iitk.ac.in

**Abstract:** The synthesis of $xAgI(1-x)[Ag_2O.B_2O_3]$ amorphous superionic conductors is done via mechanochemical synthesis route (for $x$ = 0.5 and 0.7). Ionic conductivities of $3\times10^{-3}\Omega^{-1}cm^{-1}$ for $x$ = 0.5 sample and $5\times10^{-3}\Omega^{-1}cm^{-1}$ for $x$ = 0.7 sample at room temperature, are observed, which are higher than those of the melt quenched glassy samples. Impedance plots (*Nyquist plots*) are found to be depressed semicircles with a tail at low frequency end. The corresponding equivalent circuit is postulated and each circuit element is related to a physical process. The ac conductivity $\sigma(\omega)$ is analyzed in view of the universal dynamic response (UDR), $\sigma(\omega,T) = \sigma_{dc}(T) + A_o(T)\omega^n$ (*Bode Plots*). A unique feature of the mechnochemically synthesized glasses is that their $\sigma_{dc}$ vs. 1/T behavior exhibit two distinct regions with different activation energies. The activation energy for dc conductivity ($E_{dc}$) and that for ac conductivity ($E_{ac}$) and the frequency exponent $n$ are found to satisfy the UDR–relation, $E_{ac}= (1-n)E_{dc}$ in the lower temperature regime. In the high temperature region, however, this correlation could not be established due to lack of $\sigma_{ac}$ data over the available frequency domain.

**Keywords:** Frequency response, mechanochemical synthesis, superionic glasses, ball milling.

**PACS Nos.:** 66.30Dn, 72.80Ng

## Introduction

Mechanochemical synthesis is a known and powerful technique to synthesize amorphous and nanocrystalline materials. However, only recently it has been used successfully to synthesize amorphous superionic conductors (a–SICs). Being a room temperature process, it has obvious advantages over the other conventional techniques such as rapid quenching of the melt. The work deals with mechanochemical synthesis of a–SIC and its electrical characterization by means of frequency response analysis via *Nyquist* and *Bode* plots to understand the nature of ion dynamics.

## Experimental

The starting materials were 99% pure silver iodide, silver oxide, and boron oxide supplied by Aldrich, USA. Appropriate amounts of the starting chemicals were taken in the agate pot with four balls each of diameter 18 mm and weighing 8 g. The ball is to material weight ratio was kept nearly 4 for optimum results. Milling was done in the FRITSCH Pulverisette 6 at 400 rpm in acetone medium for 30–35 hours. The acetone level was monitored periodically and kept at a constant level. Samples were taken out at regular interval to examine the progress of amorphization. After complete amorphization the residual acetone is vacuum evaporated.

XRD was done to examine amorphization of the mechanochemically synthesized material. Electrical characterization was done using HP-4192A Impedance Analyzer in the temperature range 90–425 K.

## Results and Discussion

XRD results show the formation of glass due to mechanical milling. Fig. 1 shows the XRD pattern of the sample containing 50 mol% AgI with respect to the milling time.

Nyquist plots of impedance are found to be depressed semicircles, which suggest the presence of a constant phase element (CPE) in parallel with a resistance like behavior of the system [1]. The constant phase element exponent or the depression parameter $\alpha$ is calculated for these depressed semicircles and found to be 0.48±0.04. Amount of distortion in these depressed semicircles increases with increasing temperature (Fig. 2). The pure resistance ($R_{dc}$) is calculated form the **Re**($Z^*$)–axis intercept of the depressed semicircle and used for the evaluation of dc conductivity ($\sigma_{dc}$). Fig. 3 (a) shows the equivalent electrical circuit of the cell and in Fig. 3 (b), each circuit element is related to a physical process. The CPE-1 is due to the parallel plate capacitor formed by two silver electrodes, R is due to the electrolyte resistance and CPE-2 is due to the charge double layer formed at the electrode-electrolyte interface due to polarization at lower frequencies.

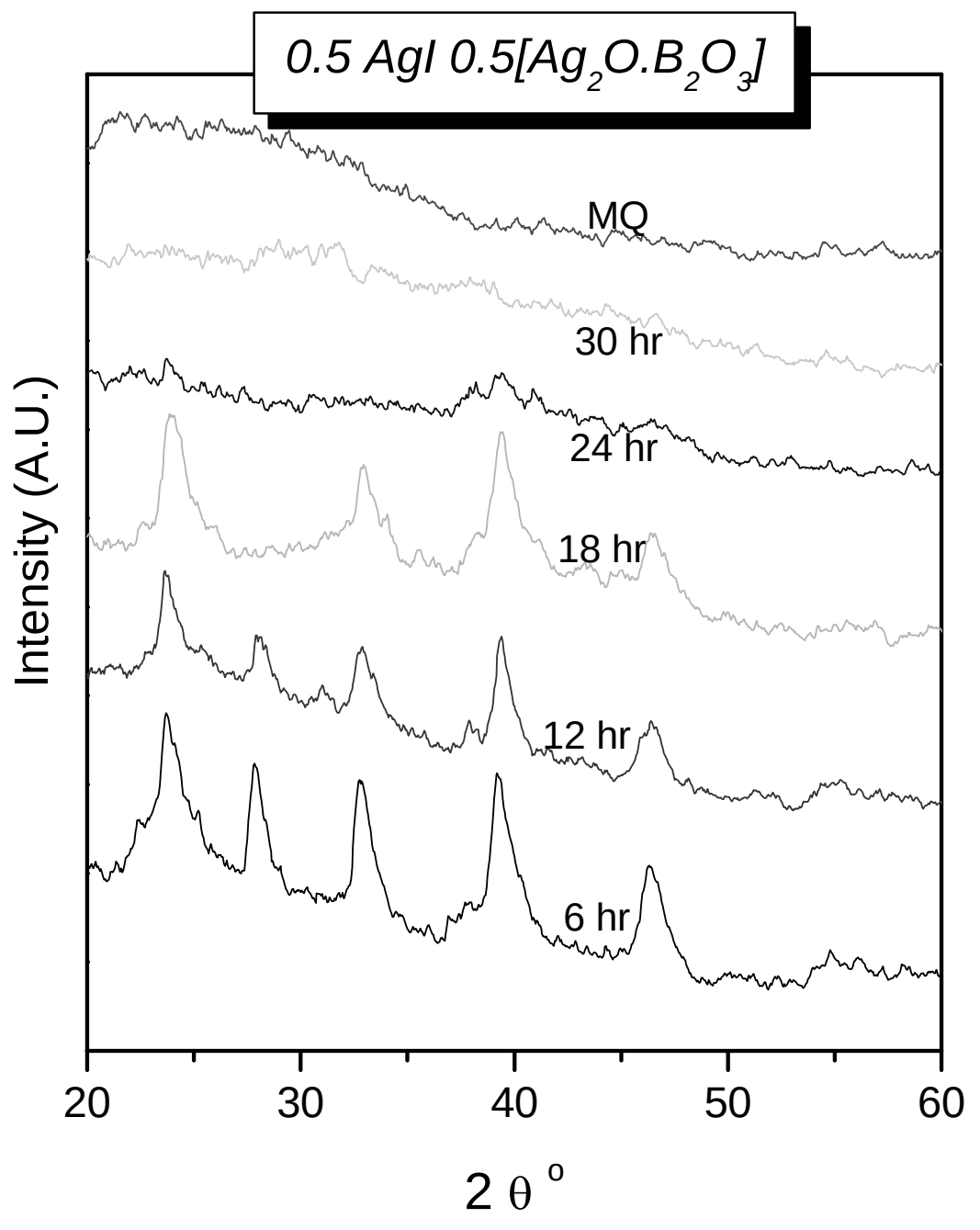

**Fig. 1: X-ray diffraction patterns of *0.5AgI-0.5($Ag_2O$-$B_2O_3$)* samples mechanically milled for different durations at room temperature. Also shown is the result for melt quenched glass of the same composition.**

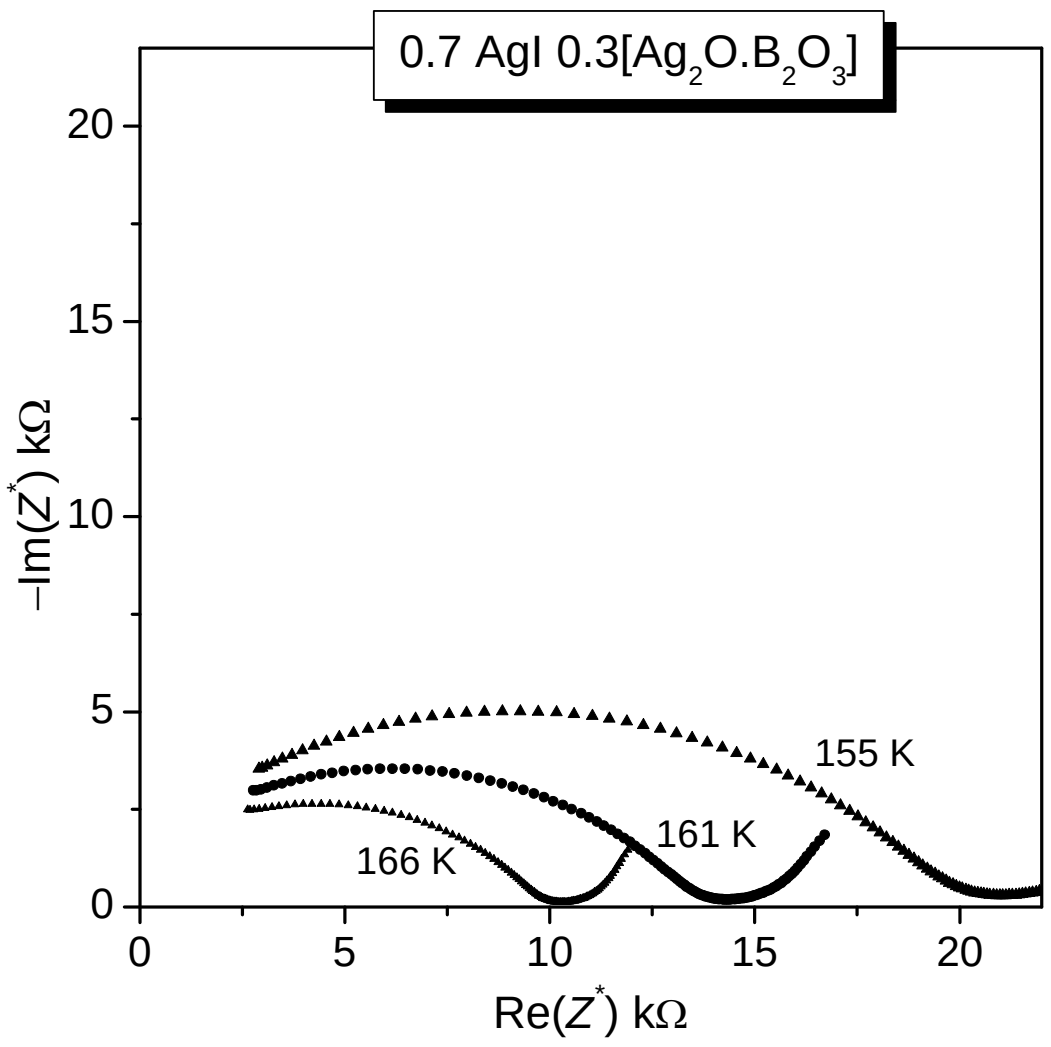

**Figure 2: Nyquist plot of impedance for 70 mol % AgI samples at 155, 161 and 166 K.**

Fig. 4 shows the electrical conductivity as a function of frequency for 50 mol% *AgI* sample in the temperatures range 124–248 K. The frequency dependence of electrical conductivity in glasses is usually described by the so called universal dynamic response [2-3].

$$\sigma(\omega,T) = \sigma_{dc}(T) + A_o(T)\omega^n \qquad \ldots \qquad (1)$$

Where $0< n \leq 1$ and $\sigma_{dc}$ is the frequency independent dc conductivity. $\sigma_{dc}$ is calculated from the plateau on the low frequency side of $\sigma(\omega)$ isotherms (Fig. 4). The other two parameters of Eq. (1), i.e., $A$ and $n$, are evaluated by allometric curve fitting [4]. At higher temperature when the glassy samples become more conducting, the electrical conductivity is observed to decrease at lower frequencies (<500 Hz), which can obviously be attributed to polarization effects. Thus the low frequency data (<500 Hz) have been neglected for the fitting purpose.

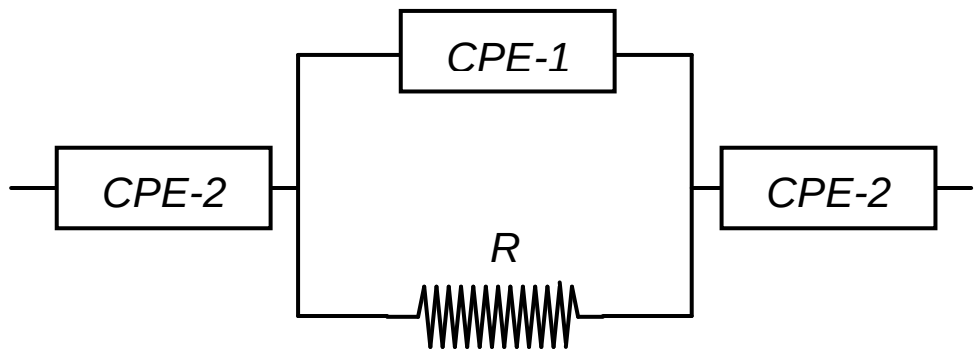

(a)

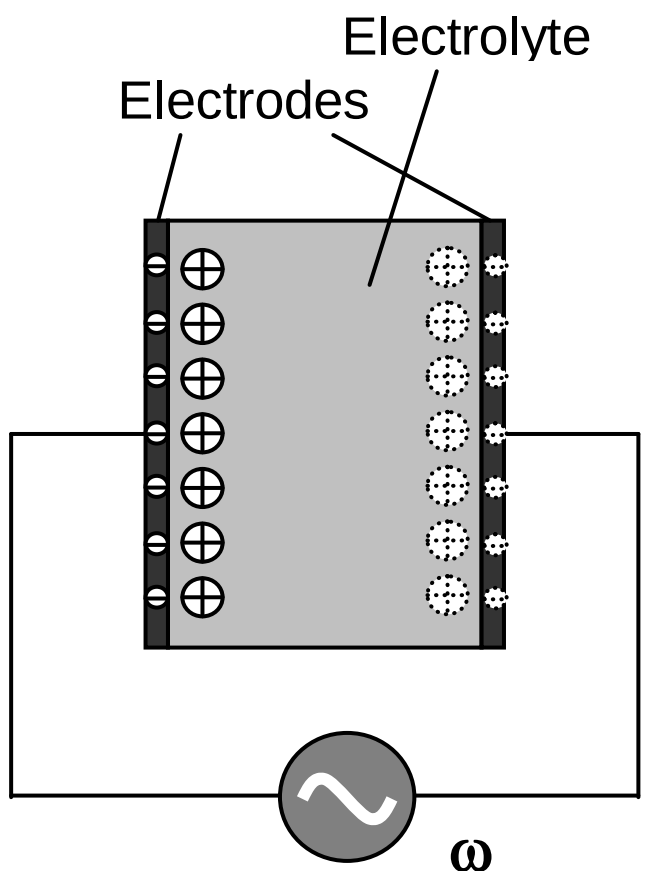

(b)

**Figure 3: (a) Equivalent circuit diagram of the cell and (b) the physical processes occurring in the cell.**

Fig. 5 compares the dc conductivities of the two samples, *i.e.*, 50 and 70 mol% *AgI*, in the whole temperature range 150–420 K. It is clearly noticed that there are two different conduction regimes, a low temperature region with a low activation energy (~0.15 eV) and a high temperature region with a higher activation energy $E_{dc2}$ [5]. While both compositions have nearly same activation energy (~0.15 eV) within the experimental error (± 0.01 eV) in the low temperature regime, the 50 mol% AgI sample has higher activation energy (0.26 eV) than the 70 mol% AgI sample in the high temperature regime. The existence of two different activation regimes is a rather unusual behavior which may suggest the presence of two distinct bonding states of $Ag^+$ ions in MM samples. It may be pointed out that MM SICs are reported to have two distinct crystallization temperatures [6], which is attributed to the possible existence of two different environments for $Ag^+$ ions.

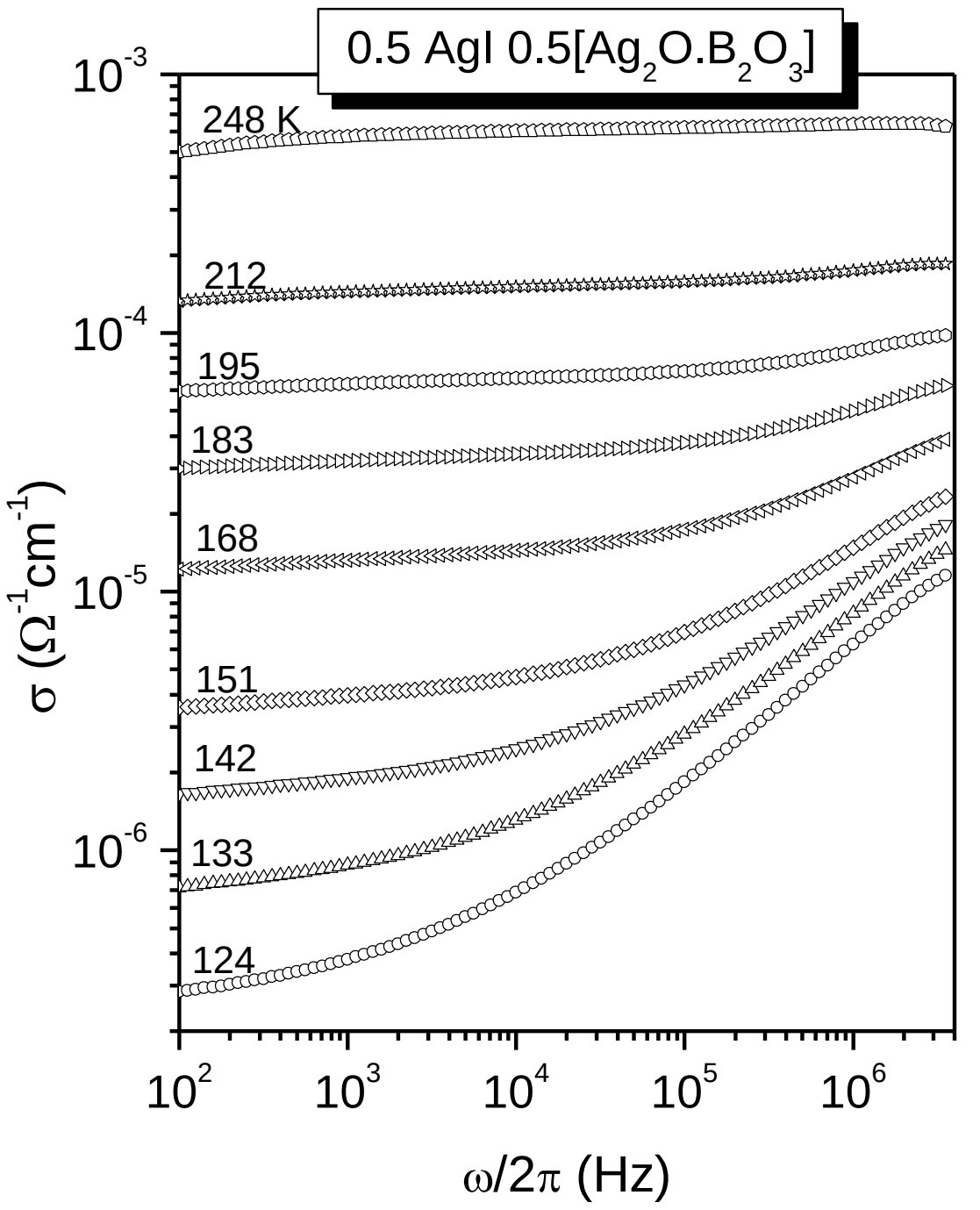

**Figure 4: Electrical conductivity as a function of frequency in the temperature range 124–248 K for 50 mol % AgI sample (Bode plot).**

The other two parameters, *i.e.*, $A$ and $n$, of the UDR (Eq.1) are evaluated by the best fit method at different temperatures. Their variation as a function of inverse temperature is shown in Fig. 6. For both compositions $A$ is found to be a thermally activated parameter with activation energies of $E_{ac}$ = 0.09 and 0.06 *eV* for 50 and 70 mol% *AgI* samples. However, the frequency exponent $n$, which is supposed to be a constant ($0<n<1$), is found to be a weakly temperature dependent parameter, decreasing with increasing temperature. The activation energies $E_{ac}$ and $E_{dc}$ and the exponent $n$ are related as

$$E_{ac} = (1-n).E_{dc} \quad \ldots \quad \ldots \quad (2)$$

This condition is satisfied by the two amorphous samples in the low temperature regime if an average value of $n$ is considered. On the basis of least square linear fit of the Figs. 4 and 5, the whole conductivity can be represented as a function of frequency and temperature as

$$\sigma(\omega,T) = \sigma_o \, \mathrm{Exp}(-E_{dc}/kT) + A_o \, \mathrm{Exp}(-E_{ac}/kT)\omega^{n(T)} \ldots (3)$$

Exponential temperature dependence of $\sigma_{dc}$ and $A$ (Eq.2) is widely accepted. But the temperature dependence of $n$ is ambiguous. Weak temperature dependence of n was reported in the introductory papers by Jonscher [2-3]. Some others have reported $n$ as an intrinsic material property and independent of temperature. The present study supports weak temperature dependence of $n$.

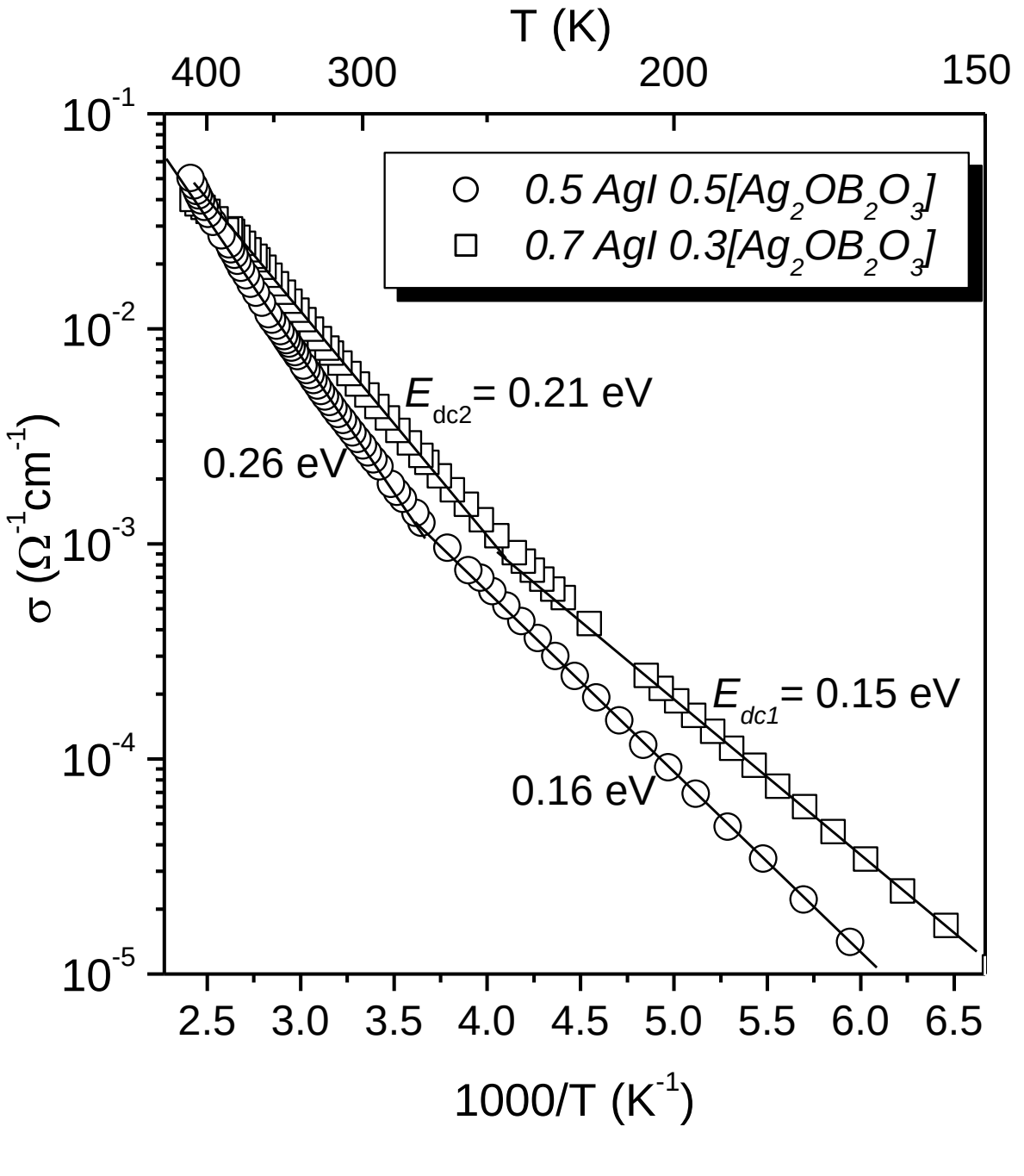

**Figure 5: DC conductivities of 50 and 70 mole% AgI samples as a function of inverse temperature. Two distinct Arrhenius regions are evident.**

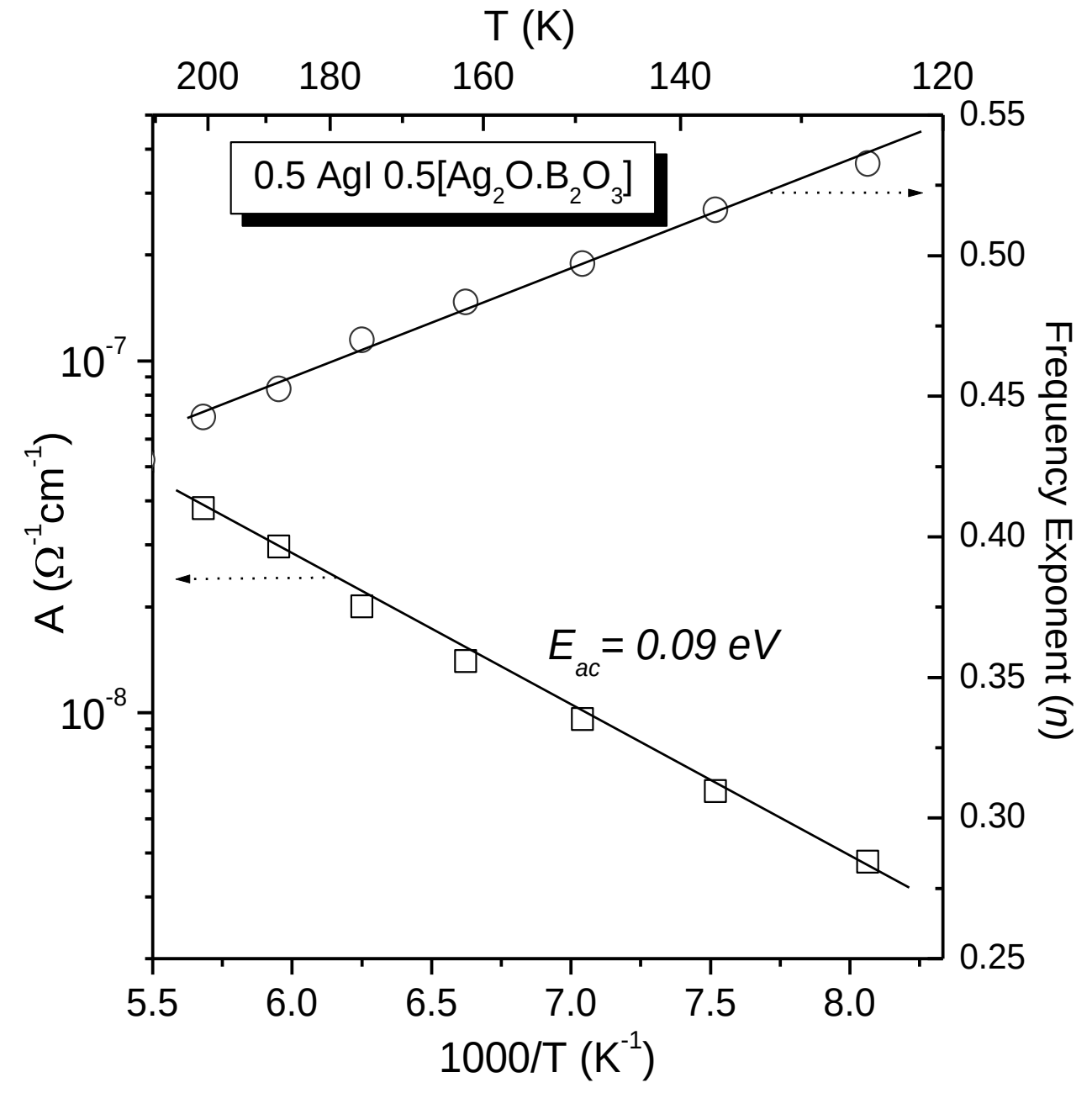

**Figure 6: The variation of parameters *A* and *n* of UDR (Eq. 1) as a function of inverse of temperature for 50 mol % AgI sample. Clearly, n is decreasing with increasing temperature.**

## Conclusion

Mechanically milled (*MM*) glasses are found to be different from the melt quenched (*MQ*) glasses in certain respects. For instance, the electrical conductivity of *MM* samples of $0.5AgI0.5[Ag_2O.B_2O_3]$ is higher ($\sigma$ ~ $3\times10^{-3}\Omega^{-1}cm^{-1}$ at 25°C) and activation energy ($E_{dc}$ ~ 0.16 *eV*) lower than that for *MQ* glasses ($\sigma$ ~ $2\times10^{-3}\Omega^{-1}cm^{-1}$, $E_a$ ~ 0.36 *eV* at 25°C). In addition to this, the $\sigma_{dc}(T)$ for *MM* samples has two distinct Arrhenius regions, while the *MQ* glasses just have one region. Also a relatively stronger temperature dependent *n* predicts some fundamental difference in the mechanism of conduction in *MM* glasses. The $yAg_2O\text{–}zB_2O_3$ glass is a poor ionic conductor ($\sigma < 10^{-5}\Omega^{-1}cm^{-1}$), so it is obvious that the mobile $Ag^+$ ions are contributed by *AgI* only. Also the ratio of $y$**:**$z$ is a defining parameter for the nature of glass. Hence, electrical measurements on $xAgI.(1\text{–}x)[yAg_2O.zB_2O_3]$ system is required for different $y$**:**$z$ ratios keeping $x$ constant to understand the change in the nature of glass with varying $y$**:**$z$ ratios. Also keeping $y$**:**$z$ ratio constant and varying $x$ one can find the fraction of $Ag^+$ ions from the *AgI* contributing to the conductivity.